\documentclass[superscriptaddress,amsmath,amssymb,aps,
prd,showkeys,
nofootinbib
]{revtex4-1}

\usepackage{varwidth}
\usepackage{multirow}
\usepackage{dcolumn}
\usepackage{tabularx}
\usepackage{booktabs}
\newcolumntype{C}{>{\centering\arraybackslash}X}

\usepackage{graphicx}
\usepackage{epstopdf}
\usepackage{graphics}
\usepackage{graphicx}
\usepackage{epsfig}
\usepackage{epstopdf}
\usepackage{makecell}
\usepackage{hyperref}
\hypersetup{
  colorlinks=true,
  linkcolor=blue,
    citecolor=blue,
    urlcolor=blue,
    anchorcolor=blue}

\begin{document}

\preprint{}

\title{Study the structure of $X(3872)$ from its lineshape}

\author{Hongge Xu}%
\author{Ning Yu}
\email{ning.yuchina@gmail.com}
\author{Zuman Zhang}
\affiliation{School of Physics and Mechanical Electrical \& Engineering, Hubei University of Education, Wuhan 430205, China}
\affiliation{Institute of Theoretical Physics, Hubei University of Education, Wuhan 430205, China}

\date{\today}

\begin{abstract}
We fit the invariant mass distribution of ${X(3872)}\rightarrow{J}/{\psi}\pi^+\pi^-$ from LHCb  using the propagator for S-wave near-threshold states in effective field theory. In this way, we can directly determine the $Z$ which measures the projection of the bound state on the compact state in ${X(3872)}$. Consequently, the structure of ${X(3872)}$ can be elucidated. Moreover, the fitting result also can describe well the data for ${X(3872)}\rightarrow{D}^{0}\overline{D}^{0*}$ from Belle experiment, which demonstrate the reliability of our fitting. The fitting indicates that $Z$ is a non-vanishing value within error, which supports that $X(3872)$ has a compact short-distant core. 
\end{abstract}

\maketitle

\section{Introduction}
A number of quarkonium-like states, which have been experimentally observed, do not fit within the conventional quarkonium spectrum, making them popular candidates for exotic hadrons. In recent decades, hadron spectroscopy has been revived by the observation of many excited heavy hadrons~\cite{ParticleDataGroup:2022pth}, the exotic $X$, $Y$, and $Z$ states~\cite{ParticleDataGroup:2022pth}, and the hidden-charm pentaquark states\cite{lhcb2015observation,lhcb2019observation}, the doubly charmed teraquark state $T_{cc}$~\cite{lhcb2022observation,lhcb2022study}, and others. The $X(3872)$ state was first discovered in the $J/ \psi\pi^{+}\pi^{-}$invariant mass spectrum from a ${B}$ meson decay by the Belle experiment~\cite{Belle:2003nnu} and has since then been confirmed by multiple experiments (BaBar~\cite{aubert2005study},CDF~\cite{acosta2004observation}, D0~\cite{abazov2004observation}, CMS~\cite{collaboration2013measurement}, LHCb experiment~\cite{gersabeck2012observation} ) from both ${B}$ meson decays and in prompt production from $p\overline{p}$ or $pp$ collisions. Since then, extensive research has been conducted to study the properties of the ${X(3872)}$ state which is probably the best known representative of these states, however, the nature of the ${X(3872)}$ state remains a puzzle ( can be see the review articles~\cite{R18} for more details). One of the peculiarities of this hadron that makes it so puzzling is that its mass is extremely fine-tuned to the ${D}^{0}\overline{D}^{0*}$ threshold. Therefore, more experimental data and theoretical developments are required to clarify the nature of these near-threshold states.

Very recently, the LHCb collaboration reports the observation of decay mode ${X(3872)}\rightarrow\psi(2S)\gamma$, and measures the ratio $R_{\psi\gamma}=\frac{\Gamma_{X(3872)\rightarrow\psi(2S)\gamma}}{\Gamma_{X(3872)\rightarrow J/\psi\gamma}}=1.67\pm0.21\pm0.12\pm0.04$~\cite{LHCb:2024tpv}. The measured ratio strongly indicates a sizeable compact component in the ${X(3872)}$ state. In Ref~\cite{Chen_2015}, an effective field theory was proposed which incorporates Weinberg's compositeness theorem and takes the compact component into account (we will call this effective field theory as CEFT in the following). The CEFT is further used to study the structure of $Z_b$~\cite{Huo:2015uka}. In this work, we will study the structure of the $X(3872)$ state by using the propagator for S-wave near-threshold state in CEFT to fit the data from LHCb and Belle collaborations. Our work is organized as follows: in section~\ref{sec2}, we briefly review the propagator for S-wave near-threshold state in CEFT. In section~\ref{sec3}, we present our numerical results; Finally, a brief summary is given in section~\ref{sec4}.

\section{The propagator for S-wave near-threshold states}\label{sec2}
In 1960s, Weinberg proposed a compositeness criterion originally employed to investigate the possibility for the deuteron to be an elementary particle, rather than a composite bound state of proton and neutron~\cite{Weinberg:1962hj,weinberg1965evidence}. The method connects the field renormalization constant ${Z}$ with effective range expansion parameters by,
\begin{equation}
a_s=[2(1-Z)/(2-Z)]/\sqrt{2\mu{B}}+O({m_\pi}^{-1}),
\end{equation}\label{eq1}
\begin{equation}
r_e=-[Z/(1-Z)]/\sqrt{2\mu{B}}+O({m_\pi}^{-1}),
\end{equation}\label{eq2}
where ${B}$ is the binding energy of deuteron and $\mu$ is the reduced mass of $n-p$. The parameters $a_s$ and $r_e$ are the scattering length and effective range, respectively. ${Z}$ measures the projection of the bound state on the bare state~\cite{Esposito_2023}. 

This compositeness criterion is being used for exotic states too. For example, LHCb collaboration presented the relative fractions of molecular and compact components in the ${X(3872)}$ state using Flatt$\acute{e}$ amplitude based on a sample of ${X(3872)}\rightarrow{J}/{\psi}\pi^+\pi^-$ candidates produced in inclusive b-hadron decays. They first search the poles in the Flatt$\acute{e}$ amplitude, and then determine $Z=15\%$  with the relation between $Z$ and the pole positions\cite{lineshapeLHCb}. Ref~\cite{esposito2022line} did a refined analysis of the LHCb experiment data and find that $0.052<Z<0.14$. BESIII collaboration carried out a similar analysis with BESIII data and obtained $Z=0.18$\cite{ablikim2023coupled}. Interestingly, LHCb collaboration further find that the lineshape can be well described both by Breit-Wigner and Flatt$\acute{e}$ amplitude. And after folding with the resolution function and adding the background, the Breit-Wigner and Flatt$\acute{e}$ lineshape are indistinguishable. Thus, in their paper they states that:``The result highlights the importance of a proper lineshape parametrization for a measurement of the location of the pole."

A propagator function different from Breit-Wigner and Flatt$\acute{e}$ is given in CEFT~\cite{Chen_2015}. A recent study further shows that the propagator proposed in CEFT is a more general formula to describe the S-wave near-threshold states\cite{Xu:2024vne}. Especially, it naturally extend Weinberg's compositeness theorem to resonances which have no-vanishing decay width. The propagator can be used whether the state is a compact state, a molecular state. In the CEFT, the propagator of the ${X(3872)}$ state (of neutral channel) can be written as\cite{Chen_2015}

\begin{equation}
G_X(E)=\frac{iZ}{E+B+\widetilde{\Sigma^\prime}(E)+i\Gamma/2},
\end{equation} \label{eq3}
where,
\begin{equation}
\widetilde{\Sigma^\prime}=-g^2[\frac{\mu}{2\pi}\sqrt{-2\mu{E}-i\epsilon}+\frac{\mu\sqrt{2\mu{B}}}{4\pi{B}}(E-B)],\label{eq4}
\end{equation}
\begin{equation}
g^2=\frac{2\pi\sqrt{2\mu{B}}}{\mu^2}(1-Z).
\end{equation}\label{eq5}
Here, $\mu$ is the reduced mass of neutral $DD$, ${E}$ is the energy defined relative to the $DD$ threshold, ${B}$ is the binding energy(we call it binding energy in the sense that it is defined relative to the threshold), $\Gamma$ can be interpreted as the renormalized decay width which comes from the non-$D\bar{D}^*$ decay of the compact state~\cite{Xu:2024vne}. 

For the $X(3872)$ state, the charged $DD$ channel may also need to be taken into account besides the neutral channel. The full propagator considering the charged $DD$ channel can be written as~\cite{Xu:2024vne}

\begin{equation}
    G_{X(3872)}(E)=\frac{iZ}{E+B+\widetilde{\Sigma^\prime}_{X(3872)}+i\Gamma/2},  \label{sec3:10}
    \end{equation}\label{eq6}
where
\begin{eqnarray}
    \widetilde{\Sigma^\prime}_{X(3872)}&=&g^2\frac{\mu}{2\pi}(\sqrt{2\mu B}-\sqrt{-2\mu{E}-i\epsilon}) \nonumber\\
    &&+g_c^2\frac{\mu_c}{2\pi}(\sqrt{2\mu_c(B+\delta)}-\sqrt{-2\mu_c(E-\delta)-i\epsilon})-(1-Z)(E+B)  \nonumber\\
    &=&-g^2[\frac{\mu}{2\pi}\sqrt{-2\mu{E}-i\epsilon}+\frac{\mu\sqrt{2\mu B}}{4\pi B}(E-B)] \nonumber\\
    &&-g_c^2[\frac{\mu_c}{2\pi}\sqrt{-2\mu_c(E-\delta)-i\epsilon}+\frac{\mu_c\sqrt{2\mu_c(B+\delta)}}{4\pi(B+\delta)}(E-B-2\delta)],  \label{charged Gx3872}
\end{eqnarray} 

\begin{equation}
    g^2\frac{\mu^2}{2\pi\sqrt{2\mu B}}+g_c^2\frac{\mu_c^2}{2\pi\sqrt{2\mu_c(B+\delta)}}=1-Z, \label{sec2:18}
\end{equation}
    
Here, $\mu_c$ is reduced mass of charged $DD$, $\delta$ is the mass splitting between the charged channel and neutral channel. In the denominator of propagator $G_X(E)$, $\Gamma= Z \Gamma_0$. The $Z$ and $\Gamma$ both are free parameters, it is not necessary to fit individually. If $Z=0$, the bound state is pure molecular state, the $\Gamma \neq 0$, which  can mimic the annihilation effect between the molecular components.

When can conclude that the propagator $G_{X}(E)$ is more general formula if comparing with the the Breit-Wigner, Flatt$\acute{e}$, low-energy amplitudes\cite{Xu:2024vne}. Firstly, for $Z=1$($g^2=0$), the propagator in CEFT reduces to Breit-Wigner. However, the Breit-Wigner amplitude assumes a compact state lies below the threshold and does not couple to the continuum state or a compact state lies above the threshold without knowing the coupling strength between this state and the continuum state. Secondly, for $Z=0$, it reduces to low energy amplitude\cite{Braaten:2007dw}. The low-energy amplitude proposed in Ref.\cite{Braaten:2007dw} can only be used for a pure molecule. So, the low-energy amplitude proposed in Ref.\cite{Braaten:2007dw} assumes the below threshold state to be a pure molecule.This point has already been mentioned in Ref.\cite{Chen_2015}. Finally, for $0<Z<1$, it is equivalent to Flatt$\acute{e}$ amplitude\cite{Flatte:1976xu}. It can be observed that as $Z$ approaches 0, the Flatt$\acute{e}$ parameters ($E_f,g_1$) become infinite, due to ($B_0,g_0$) becoming infinite. Conversely, as $Z$ approaches 1, $g_1$ becomes 0, coinciding with the vanishing of $g_0^2$. Hence, the Flatt$\acute{e}$ parameterization can only be successfully applied in the case $0<Z<1$, in the other word, the Flatt$\acute{e}$ amplitude assumes the existence of a compact object. 
In summary, all the three amplitudes have assumptions. In contrast, the propagator $G_X(E)$ derived from CEFT includes the factor $Z$ explicitly, thus it makes no assumptions on the structure of the near-threshold states and provides a more general formula to describe S-wave threshold states. 

In this work, we will use the propagator without and with considering charged channel to fit the lineshape of the ${X(3872)}$ state from LHCb experiment, respectively. In this way, we can determine ${Z}$ directly from the fitting.
    We would like to compare our fitting scheme with Ref\cite{Chen_2015} before ending this section. In Ref\cite{Chen_2015}, two mechanisms are considered in the production of the ${X(3872)}$ state. One is that the ${X(3872)}$ state is produced directly in the short-distant vertex. The other is that a ${D}^{0*}\overline{D}^{0}$ pair is produced first and then rescatters into the ${X(3872)}$ state. The first one is called as short-distance production mechanism, and the second one is called as long-distance production mechanism in Ref\cite{Chen_2015}. The short-distance production mechanism amplitude (denoted as $i\mathcal{M}_s$) is attributed to the compact component of the ${X(3872)}$ state, while the long-distance production amplitude (denoted as $i\mathcal{M}_l$) is attributed to the ${D}^{*0}\overline{D}^{0}$ component of the ${X(3872)}$ state. It is found in Ref\cite{Chen_2015}, $i\mathcal{M}_s\sim G_X$ and is $\mathcal{O}(p^{-3/2})$, where $p$ is the small momentum scale in the effective field theory, $i\mathcal{M}_l$ is of the form (see Ref\cite{Chen_2015} for more details).
\begin{equation}
i\mathcal{M}_l\sim\frac{ZE+(2-Z)B+i\Gamma/2}{E+B+\widetilde{\Sigma^\prime}(E)+i\Gamma/2},\label{eq8}
\end{equation}
and with $\mathcal{O}(p^0)$. Thus, from the power counting of effective field theory, one can find that $i\mathcal{M}_s\gg i\mathcal{M}_l$. This is confirmed by the numerical study of Ref\cite{Chen_2015} (see fig3 there), where it is shown that $i\mathcal{M}_l$ gives almost negligible contribution to the lineshape. Ref\cite{Chen_2015} then concludes that if $Z$ is non-negligible, the lineshape will be driven by the short-distance production mechanism, even if the dominant component of the ${X(3872)}$ state is ${D}^{*0}\overline{D}^{0}$. Therefore, if the ${X(3872)}$ state contains compact component, it is appropriate to simply use the propagator $G_X$ to fit the lineshape. However, if the ${X(3872)}$ state is a pure molecular state, only the long-distance production mechanism contributes to the lineshape, one may wonder whether it is appropriate to simply use the $G_X$ to fit the lineshape in such case. This answer is evident by noting  that if $Z=0$ the energy-dependent term $ZE$ of the numerator in Eq.\eqref{eq8} disappears. Consequently, the energy dependent of the lineshape can also be simply described by the propagator $G_X$, if the ${X(3872)}$ state is a pure molecular state. Therefore, we have demonstrated that using  $G_X$ in CEFT to fit the lineshape is appropriate regardless of whether the $X(3872)$ state is a pure molecular state or contains compact component.

\section{Result}\label{sec3}
The nature of the ${X(3872)}$ state can be elucidated by studying its lineshape. LHCb collaboration analyzed a sample of ${X(3872)}\rightarrow{J}/{\psi}\pi^+\pi^-$ candidates in  ${P}_{\pi^+\pi^-}\leqslant 12$ GeV, $12\leqslant{P}_{\pi^+\pi^-}\leqslant 20$ GeV,~and $20\leqslant{P}_{\pi^+\pi^-}\leqslant 50$ GeV collected in 2011 and 2012 to determine the lineshape of the ${X(3872)}$ state\cite{lineshapeLHCb}. Recently, Belle collaboration also measured the lineshape of the ${X(3872)}$ state based on a data sample in the decay $B\rightarrow{X(3872)}K\rightarrow{D}^{0}\overline{D}^{0*}K$\cite{lineshapeBelle}. The data of LHCb experiment have higher statistical and smaller errors compared to Belle experiment, enabling more accurate fitting parameters of the ${X(3872)}$ state. The main purpose of this work is to fit the parameter of the ${X(3872)}$ state using data from LHCb experiment, the fitting result based on the data of Belle experiment will be treated as a comparison. 
\begin{figure}[htbp] 
    \includegraphics[width=0.45\textwidth]{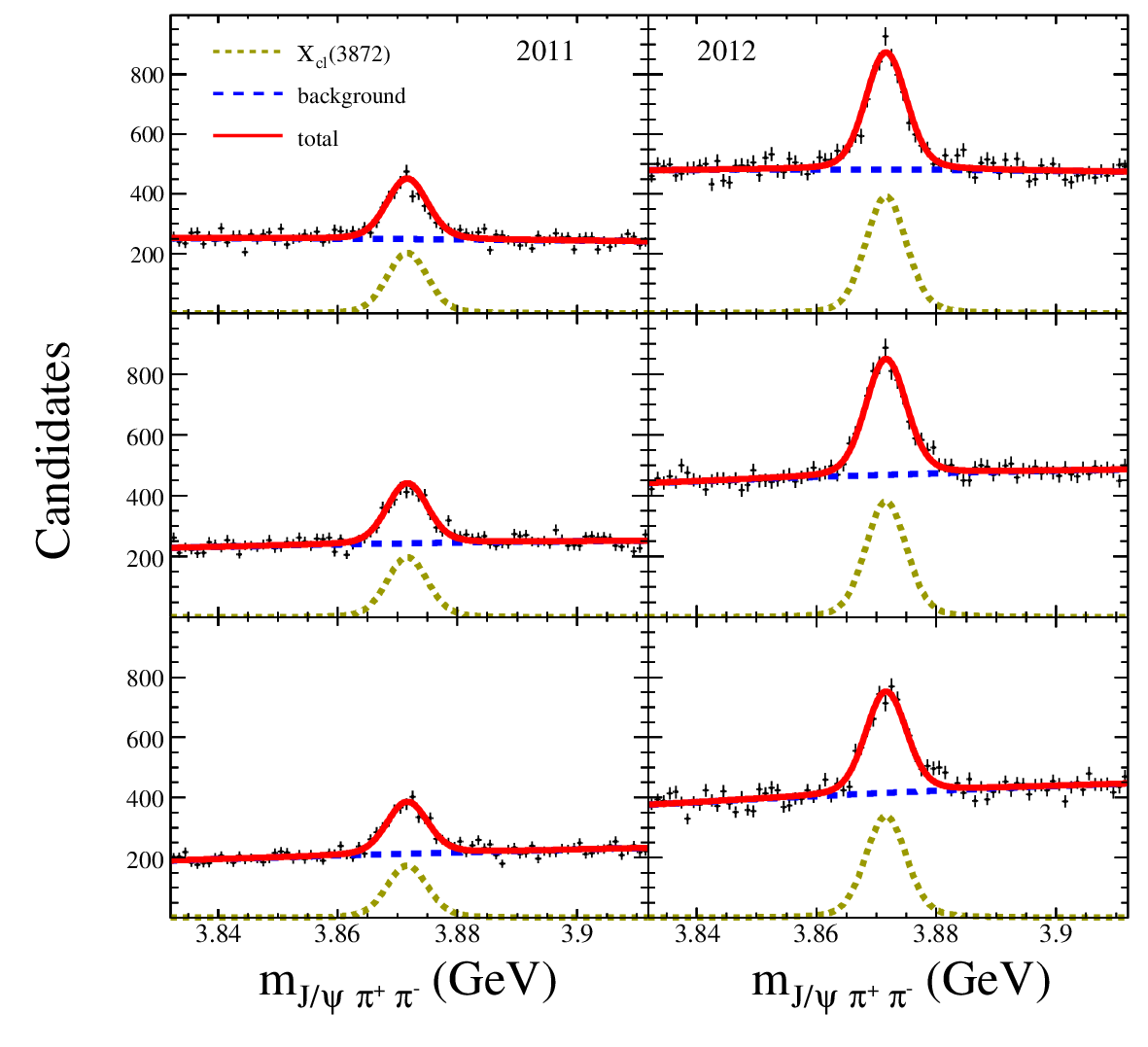}
    \caption{Mass distributions of the $X(3872)$ state based on LHCb experiment data. The distribution in top, middle, and  bottom is performed in ${P}_{\pi^+\pi^-}\leqslant 12$ GeV, $12\leqslant{P}_{\pi^+\pi^-}\leqslant 20$ GeV,$20\leqslant{P}_{\pi^+\pi^-}\leqslant 50$ GeV, respectively. The left- (right-) hand plot is for 2011 (2012) data. The points with error bars represent data. The red solid line shows the total fit result. The blue dashed line shows the contribution of generic background. In the distributions the charged channel is taken into account.}
    \label{tu2}
    \end{figure}


For the ${X(3872)}$ state, $B$, $\Gamma$ and $Z$ these parameters can be determined by fitting the lineshape data using its propagator. When we only consider neutral channel, we find that the result is $Z=0.42\pm0.16$ using the data from LHCb experiment. While the charged $DD$ channel is also taken into account, these parameters are a little different, but $Z$ also is a non-vanished value and they are consist with each other within error. These free parameters result are shown in Table~\ref{paci1}. The mass distribution of the ${X(3872)}$ state for the decay channel ${X(3872)}\rightarrow{J}/{\psi}\pi^+\pi^-$ is shown in Fig.~\ref{tu2}, with the parameters determined by fitting LHCb experiment data. In Fig.~\ref{tu2}, we considered the charge channel. When only the neutral channel is accounted for, the mass distribution of the ${X(3872)}$ state remains similar. For the analysis that includes the charge channel, compared to  the analysis based on the extrapolation to single channel case by the LHCb collaboration, our approach can directly determine the free parameters using the propagator of couple channel. 

\begin{table}[!htbp]
    \setlength{\tabcolsep}{1.5mm}
    \caption{The parameters from fitting lineshape of the ${X(3872)}$ state.}
    \begin{tabular}{cccccc} \hline  \hline 
      Fitting scheme&${Z}$ &$\Gamma$(MeV) &B(MeV) &$\chi^2$/ndf \\ \hline
      only neutral channel   &$0.42\pm0.16$  &$0.57\pm0.23$  &$0.19\pm0.05$  &76.7/77  \\
      With charged channel   &$0.49\pm0.27$  &$0.78\pm0.40$  &$0.18\pm0.06$  &79.3/77  \\     
     \hline\hline
    \end{tabular} \label{paci1}
    \end{table}

In our fitting, the resolution is modeled with a Gaussian function. We first fix the mass and width of the ${X(3872)}$ state with Breit-Wigner fitting results of LHCb experiment and setting $Z=1$ in our fitting program (CEFT propagator reduces to Breit-Wigner, as for $Z=1$). Then, we fit the width $\sigma$ of the Gaussian function and set it to be the fixed value of the $\sigma$ fixed, while $B$, $\Gamma$ and $Z$ are set as free parameters. By the way, we use one fixed $\sigma$ in all bins rather than individual $\sigma$ in each momentum bin for simplicity. In the fitting analysis that considers charge channel, we assume the coupling constants $g$ and $g_c$ in Eq.\ref{charged Gx3872} are equal.

        \begin{figure}[htbp]
            \includegraphics[width=0.45\textwidth]{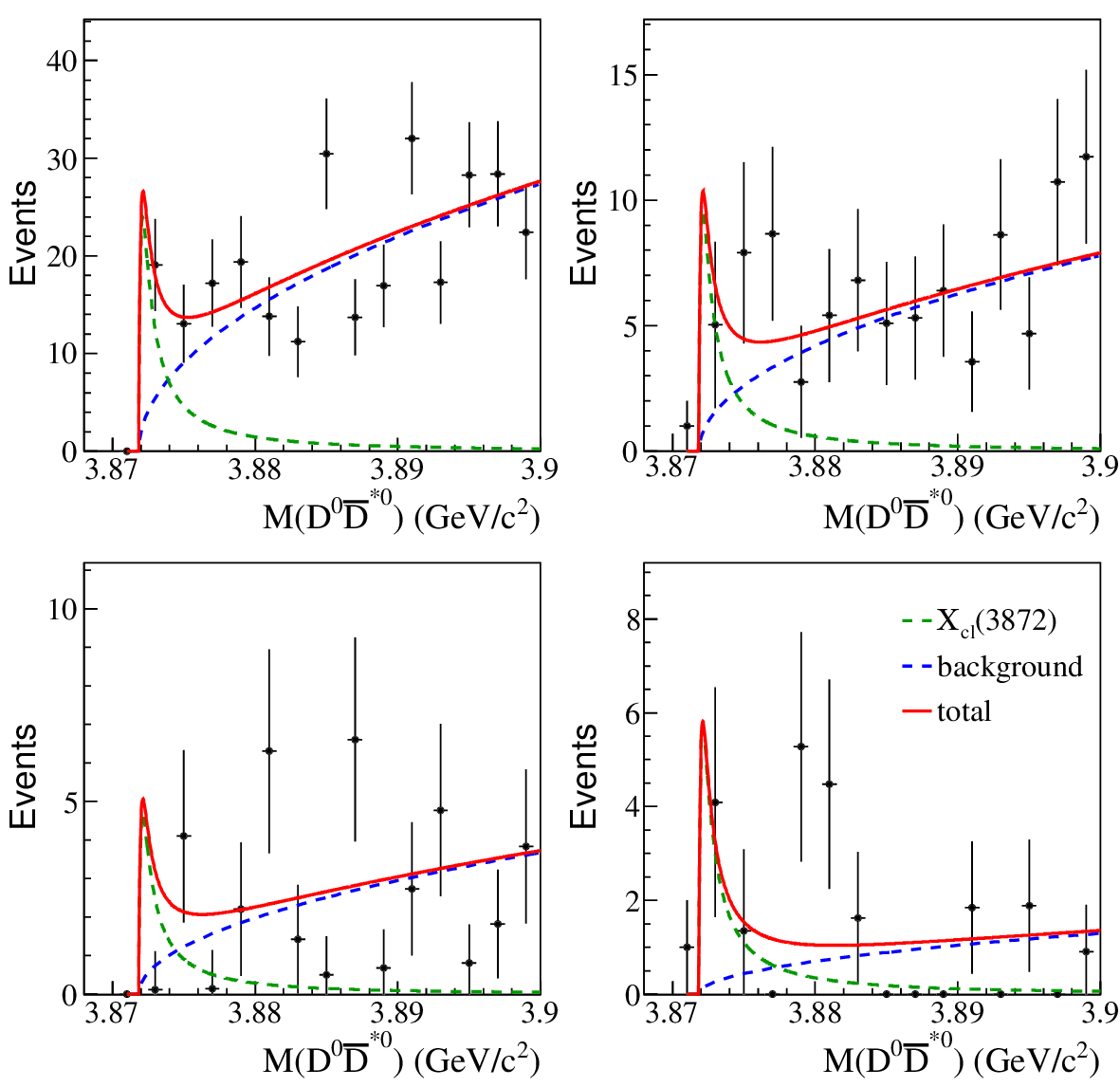}
            \caption{The $M({D}^{0}\overline{D}^{0*})$ distributions for $B^+\rightarrow{X(3872)}K^+$(top) and $B^0\rightarrow{X(3872)}K^0$(bottom). The left and right rows are for  ${D}^{0*}\rightarrow\overline{D}^{0}\gamma$ and ${D}^{0*}\rightarrow\overline{D}^{0}\pi^0$, respectively. The points with error bars represent data which have subtracted broken signal. The red solid line shows the total fit result. The green dashed lines show the contribution from the signal of the ${X(3872)}$ state. The blue dashed lines show the contribution of generic background. The data points have the broken signal remained.}
            \label{tu3}
            \end{figure}

The mass distribution of the ${X(3872)}$ state for $B\rightarrow{X(3872)}K\rightarrow{D}^{0}\overline{D}^{0*}K$ based on the data from Belle experiment, with the parameters determined by LHCb's fitting, are shown in Fig~\ref{tu3}. From the Fig~\ref{tu3}, we can conclude that the non-vanishing parameter $Z$ determined by fitting LHCb experiment data also can describe well the Belle experiment data. By above fitting result, we can find the that $Z$ is a non-vanishing value within error, which supports that $X(3872)$ has a compact short-distant core.

\section{Summary}\label{sec4}
In the work, we fit the mass distribution of the $X(3872)$ state for ${X(3872)}\rightarrow{J}/{\psi}\pi^+\pi^-$ measured by LHCb experiment and $B\rightarrow{X(3872)}K\rightarrow{D}^{0}\overline{D}^{0*}K$ measured by Belle experiment to extract the ${Z}$. We find that the ${Z}$ for the $X(3872)$ state is a non-vanishing value within error and is larger than previous results\cite{Chen_2015,lineshapeLHCb,esposito2022line,ablikim2023coupled}. The extracted ${Z}$ may indicate that the $X(3872)$ state contains a sizeable tetraquark or $c\bar c$ core which is consist with the recent discovery by LHCb collaboration~\cite{LHCb:2024tpv}. This compact core can naturally explain the large production rates~\cite{Bignamini:2009sk,Duan:2024zuo} (also the large and negative effective range, see Ref~\cite{esposito2022line,Esposito_2023}) and decay branch ratio~\cite{Grinstein:2024rcu,LHCb:2024tpv} of the $X(3872)$ state. The structure of other S-wave ${XYZ}$ states can also be analyzed using the CEFT which is in progress.

\section{Acknowledgments}
We would like to thank Guoying Chen for drawing our attention to the EFT which incorporates compositeness and valuable discussions.

\bibliography{ref}
\end{document}